\documentclass[aip, ltp,
 amsmath,amssymb,
 reprint,%
]{revtex4-1}

\usepackage{listings}
\usepackage{amsmath}
\usepackage{amssymb}
\usepackage{amsxtra}
\usepackage{physics}
\usepackage{mathrsfs}
\usepackage{booktabs}
\usepackage{gensymb}
\usepackage{graphicx}
\graphicspath{ {./images/} }
\usepackage{tikz}
\usepackage{hyperref}
\usepackage{lipsum}
\usepackage{comment}

\usepackage{mathtools}

\newcommand{\blue}[1]{{\color{black}{#1}}}

\begin{document}

\title{Subluminal and superluminal velocities of free-space photons}

\author{Konstantin Y. Bliokh}
\affiliation{Donostia International Physics Center (DIPC), Donostia-San Sebasti\'{a}n 20018, Spain}
\affiliation{IKERBASQUE, Basque Foundation for Science, Bilbao 48009, Spain}

\begin{abstract}
We consider rectilinear free-space propagation of electromagnetic wavepackets using electromagnetic field theory, scalar wavepacket propagation, and quantum-mechanical formalism. We demonstrate that spatially localized wavepackets are inherently characterized by a subluminal group velocity and a superluminal phase velocity, whose product equals $c^2$. These velocities are also known as the `energy' and `momentum' velocities, introduced by K. Milton and J. Schwinger. We illustrate general conclusions by explicit calculations for Gaussian and higher-order beams and wavepackets, and also highlight subtleties of the quantum-mechanical description based on the `photon wavefunction'. 
\end{abstract}

\maketitle 

\section{Introduction}

Subluminal and superluminal velocities in the propagation of light have intrigued scientists for several decades \cite{Brillouin_book}. 
Multiple distinct velocities characterize light propagation, such as phase, group, signal, and energy-transfer velocities. Furthermore, various physical mechanisms can lead to subluminal or superluminal behavior: interactions with matter \cite{Chiao1997, Winful2006, Boyd2002, Khurgin2010, Asano2016}, nontrivial quantum-vacuum effects \cite{Latorre1995}, local phase gradients in structured free-space light \cite{Berry2010, Berry2012, Bliokh2013NJP_II}, specially engineered space-time wavepackets \cite{Bliokh2012, Kondakci2019, Yessenov2022}, and the tilt of plane-wave Fourier components in transversely confined light beams \cite{Horvath1996APB, Porras2002PRE, Porras2003PRE, Giovannini2015, Bouchard2016O, Gouesbet2016, Bareza2016SR, Alfano2016OC, Saari2018PRA, Petrov2019SR, Bliokh2023JPA, Gouesbet2026JQSRT}. 

\begin{figure}[t]
\centering
\includegraphics[width=\linewidth]{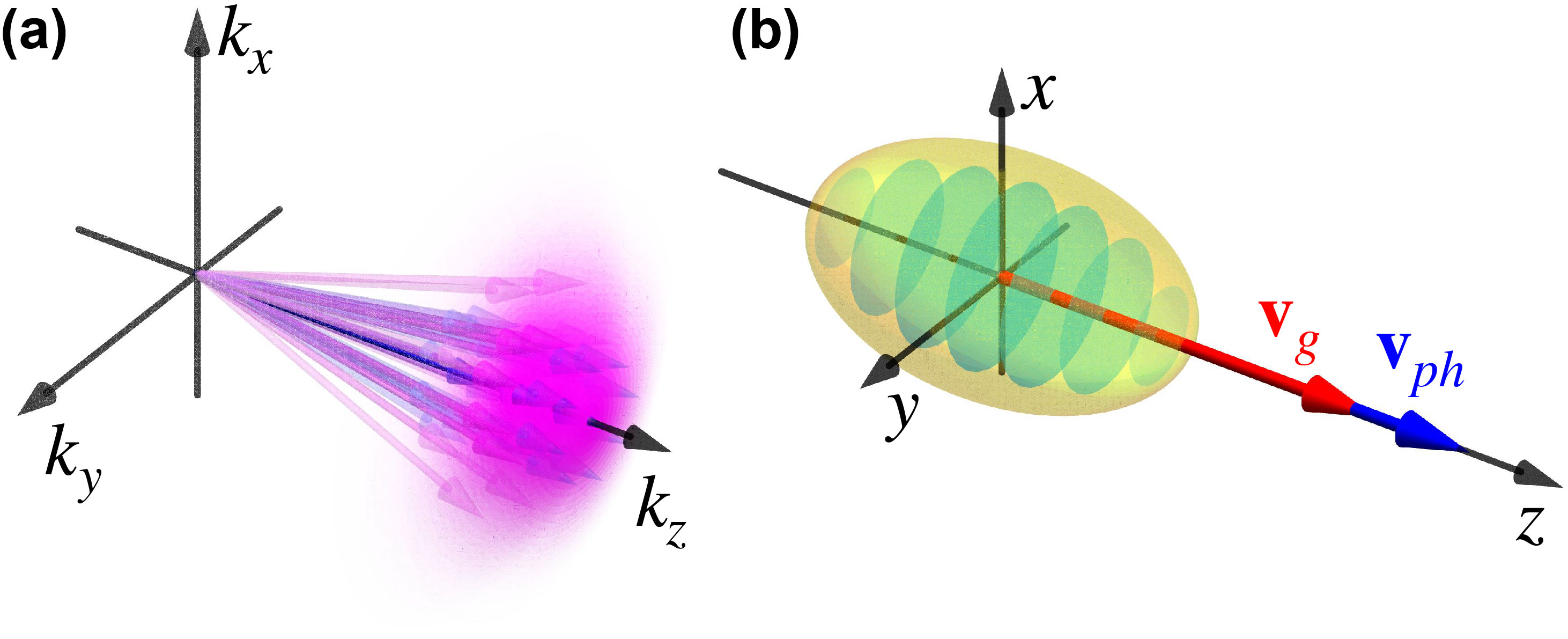}
\caption{Schematics of a localized electromagnetic wavepacket propagating along the $z$-axis: (a) its plane-wave spectrum (momentum-space distribution) and (b) its real-space picture. The wavepacket is characterized by a subluminal group velocity $v_{g} < c $ and a superluminal phase velocity  $v_{ph} > c $, such that $v_g v_{ph} = c^2$.}
\label{fig:intro}
\end{figure}

In this work, we address the simplest scenario among these, namely, the free-space propagation of electromagnetic Gaussian-like wavepackets without engineered internal structure. The only requirement is suitable spatial confinement of the wavepacket, which necessarily entails a corresponding spread in momentum (wavevector) space. As theoretically described and experimentally observed in \cite{Giovannini2015, Gouesbet2016, Alfano2016OC, Bouchard2016O, Bareza2016SR, Saari2018PRA, Bliokh2023JPA}, such wavepackets or single photons propagate with a subluminal group velocity $v_g < c$, determined by the transverse confinement of the wavepacket. 
In addition, the wavepacket can also be characterized by a superluminal phase velocity $v_{ph} > c$, such that $v_g v_{ph} = c^2$, see Fig.~\ref{fig:intro}.

We trace the origin of this result through several complementary approaches and establish useful connections between relativistic field theory, quantum mechanics, and previous studies of this phenomenon. We also examine the extension of the main results to higher-order modes, such as Laguerre-Gaussian beams \cite{Bouchard2016O, Bareza2016SR}. 

\section{Electromagnetic field-theory approach}

We begin with a textbook exposition of electromagnetism based on relativistic field theory \cite{Jackson, Soper, Schwinger_book}. It is known that an electromagnetic field can be characterized by its energy density $U$ and momentum density ${\bf P}$ (proportional to the Poynting vector). For a localized wavepacket propagating in free space, the spatial integrals of these quantities are conserved. According to Noether’s theorem, the conservation of energy and momentum follows from invariance under time and space translations, respectively. 
In addition, there are conservation laws for angular momentum and the so-called `boost momentum', associated with invariance under spatial rotations and and spatio-temporal rotations (i.e., Lorentz boosts).

Wave-packet propagation is closely related to boost-momentum conservation \cite{Schwinger_book, Bliokh2023JPA}, which can be written as
\begin{equation}
\label{boost_momentum}
\int ({\bf r} U - c^2 t {\bf P}) \,d^3{\bf r} = {\bf const}\,.
\end{equation}
From here, one readily finds that the energy centroid, ${\bf R}_E = \dfrac{\int {\bf r} U  \,d^3{\bf r}}{\int U  \,d^3{\bf r}}$, obeys the equation of motion
\begin{equation}
\label{energy_velocity}
\frac{d{\bf R}_E}{dt} 
= c^2\frac{\int {\bf P} \,d^3{\bf r}}{\int U  \,d^3{\bf r}} \equiv {\bf v}_E\,,
\end{equation}
where we have used the fact that the total energy $\int U \, d^3{\bf r}$ is a conserved time-independent quantity.

Equation \eqref{energy_velocity} represents the field-theoretical analogue of the relativistic equation of motion for a point particle,: ${\bf v} = c^2 {\bf p}/E$ \cite{LLfield}. Importantly, the energy-centroid velocity \eqref{energy_velocity} is {\it subluminal} for localized wavepackets. 
This can be seen directly from the explicit expressions for the electromagnetic energy and momentum densities: $U = ({\bf E}^2 + {\bf H}^2)/2$ and ${\bf P} = ({\bf E}\times{\bf H})/c$, where ${\bf E}$ and ${\bf H}$ are the electric and magnetic fields, and we use Gaussian units omitting inessential numerical constants. Obviously, $c|{\bf P}| \leq U$, and equality is attained for null fields with ${\bf E}\cdot {\bf H} =0$. Therefore, $|{\bf v}_E| \leq c$, where equality is possible only for null fields with a fixed direction of ${\bf P}$. This means that $|{\bf v}_E| = c$ only for plane waves, which are spatially delocalized. 

By analogy with `energy velocity' \eqref{energy_velocity}, Milton and Schwinger introduced the concept of a `momentum velocity' in their textbook \cite{Schwinger_book}. Using an electromagnetic analogue of the virial theorem, they derived the relation
\begin{equation}
\label{momentum_centroid}
\frac{d}{dt}{\int {\bf r} \cdot {\bf P}  \,d^3{\bf r}}
= {\int U  \,d^3{\bf r}} \,.
\end{equation}
This motivated the definition of the `momentum centroid' projected onto the integral momentum direction: ${\bf R}_P \cdot \int {\bf P}\, d^3{\bf r} = {\int {\bf r} \cdot {\bf P}  \,d^3{\bf r}}$. Taking the time derivative, and  using Eq.~\eqref{momentum_centroid} and the conservation of the total momentum $\int {\bf P}  \,d^3{\bf r}$, we find that this centroid propagates with a velocity ${\bf v}_P$ satisfying
\begin{equation}
\label{momentum_velocity}
{\bf v}_P \cdot \int {\bf P}\, d^3{\bf r} 
= {\int U  \,d^3{\bf r}} \,.
\end{equation}
%
\blue{In what follows, we assume that this velocity is directed along the total momentum, so it can also be written as 
\[
{\bf v}_P = \frac{\int {\bf P}\, d^3{\bf r}\,{\int U  \,d^3{\bf r}}}{|\int {\bf P}\, d^3{\bf r}|^2}\,.
\]
Comparing Eqs.~\eqref{energy_velocity} and \eqref{momentum_velocity}, we find that ${\bf v}_P \cdot {\bf v}_E =c^2$. Consequently, the `momentum velocity' is {\it superluminal} for localized wavepackets: $|{\bf v}_{P}| = 1/|{\bf v}_{E}| >c$.}

Remarkably, Milton and Schwinger concluded from Eqs.~\eqref{boost_momentum}--\eqref{momentum_velocity} that \cite{Schwinger_book}: ``{\it If the flow of energy and momentum takes place in a single direction, it would be reasonable to expect that these mechanical properties are being transported with a common velocity ${\bf v}_E = {\bf v}_P = {\bf v}$, $|{\bf v}|=c$, ..., which results express the mechanical properties of a localized electromagnetic pulse carrying both energy and momentum at the speed of light, in the direction of the momentum}''. However, this argument contains an intrinsic inconsistency: in a localized electromagnetic wavepacket the flow of energy does not take place in a single direction. Indeed, 
a wavepacket arises from the interference of multiple plane waves propagating in different directions; it diffracts during propagation, and energy flows in both longitudinal and transverse directions \cite{Ghosh2024JOSA}. (Note that non-diffracting Bessel beams \cite{McGloin2005CP} are not square-integrable and therefore do not represent properly localized fields.)

\section{Classical wavepacket considerations}
\label{sec:wavepacket}

We now consider a classical wavepacket composed of multiple plane waves with different wavevectors and frequencies satisfying the dispersion relation $\omega ({\bf k}) = ck$. For simplicity, we treat a scalar wavepacket described by the wavefunction $\psi({\bf r},t)$ and its plane-wave (Fourier) components $\tilde{\psi}({\bf k})e^{-i \omega({\bf k}) t}$. In the Fourier (momentum) representation, energy, momentum, and position correspond to the frequency $\omega$, wavevector ${\bf k}$, and the operator $i{\boldsymbol{\nabla}}_{\bf k}$, respectively. Accordingly, the energy centroid can be written as
\begin{equation}
\label{energy_centroid}
{\bf R}_E
= \frac{\int \omega\, \tilde{\psi}^* e^{i\omega t} (i{\boldsymbol{\nabla}}_{\bf k}) \tilde{\psi}e^{-i\omega t}\,d^3{\bf k}}{\int \omega |\tilde{\psi}|^2  \,d^3{\bf k}}\,.
\end{equation}
Taking the time derivative of this expression yields the energy-centroid velocity:
\begin{equation}
\label{group_velocity}
{\bf v}_E
= \frac{\int \omega\, {\bf v}_g |\tilde{\psi}|^2 \,d^3{\bf k}}{\int \omega |\tilde{\psi}|^2  \,d^3{\bf k}}
= c^2\frac{\int {\bf k} |\tilde{\psi}|^2 \,d^3{\bf k}}{\int \omega |\tilde{\psi}|^2  \,d^3{\bf k}} 
\equiv \langle {\bf v}_g \rangle \,,
\end{equation}
where ${\bf v}_g = \partial \omega /\partial {\bf k} = c{\bf k}/k$ is the local group velocity in momentum space.
\blue{(Considering the {\it $z$-propagation} of a skew plane wave, this group velocity yields $v_{g\,z} = ck_z/k <c$, while the corresponding phase velocity is $v_{ph\,z} = \omega/k_z = ck/k_z>c$.)}
Equation~\eqref{group_velocity} shows that the energy-centroid velocity can be regarded as the average {\it group velocity} of the wavepacket, and the second equality represents the momentum-representation counterpart of Eq.~\eqref{energy_velocity}. 

In turn, the average {\it phase velocity} of the wavepacket can be defined as the ratio of the mean frequency to the mean wavevector component along the propagation direction. This can be expressed as
\begin{equation}
\label{phase_velocity}
\langle {\bf v}_{ph} \rangle \cdot {\int {\bf k} |\tilde{\psi}|^2 \,d^3{\bf k}} = \int \omega |\tilde{\psi}|^2  \,d^3{\bf k}\,.
\end{equation}
This relation is fully analogous to Eq.~\eqref{momentum_velocity} for the `momentum velocity', so that  
$\langle {\bf v}_{ph} \rangle \cdot \langle {\bf v}_{g} \rangle = c^2$ and $\langle {\bf v}_{ph} \rangle = {\bf v}_P$.

Let us perform explicit calculations for paraxial wavepackets propagating along the $z$-axis. In fact, the deviations of the group and phase velocities from $c$ are governed by the {\it transverse} confinement of the wavepacket \cite{Bliokh2023JPA}, and 
it suffices to consider monochromatic beams localized only in the transverse directions. 

\begin{figure}[t]
\centering
\includegraphics[width=\linewidth]{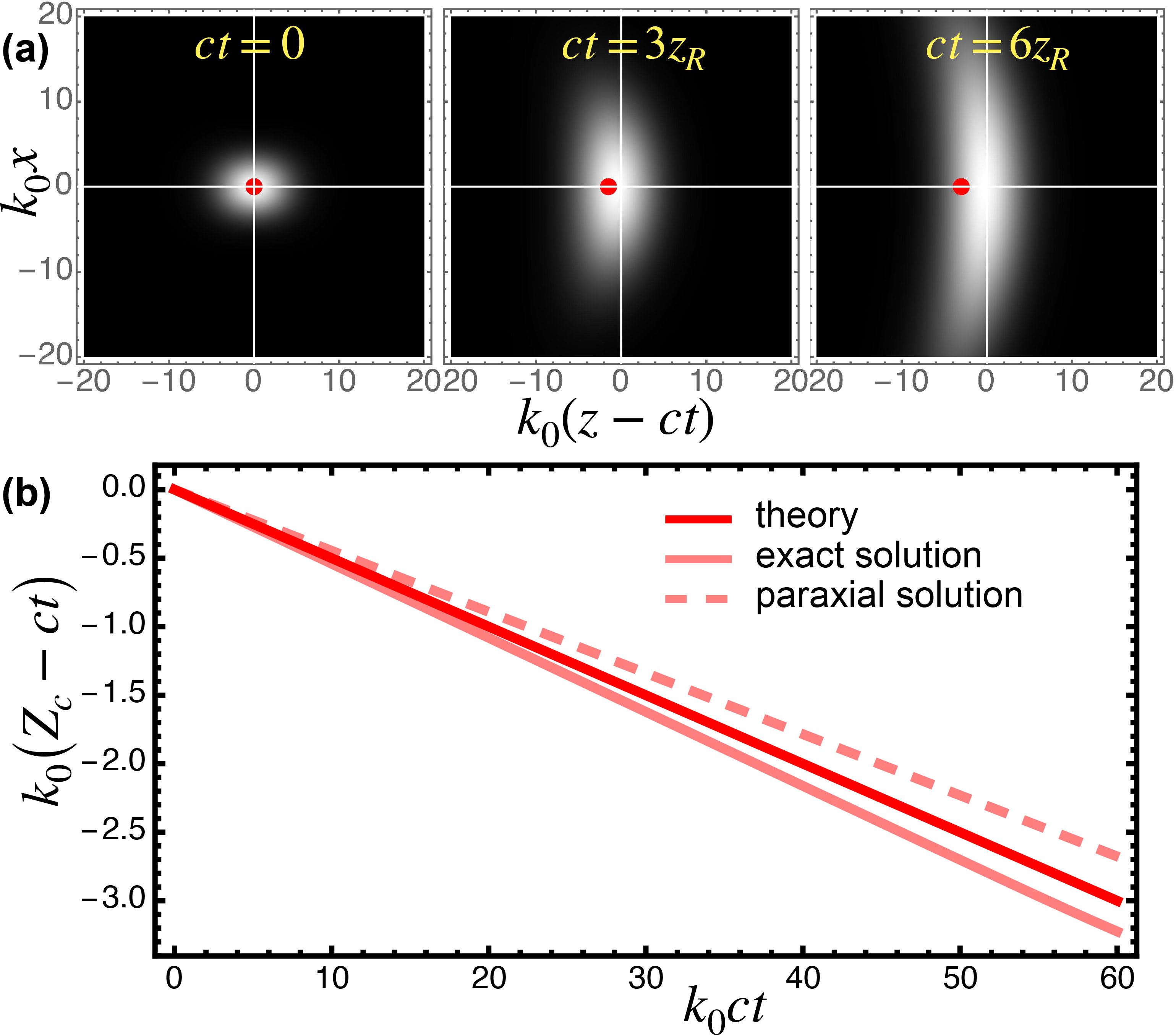}
\caption{(a) Propagation of a Gaussian-like wavepacket obtained from the exact solution of the scalar wave equation \cite{Vo2024JO} [Eq.~(3) therein], constructed as a superposition of Gaussian-like beams with a common Rayleigh range $z_R$ and frequencies distributed according to a Poisson-like spectrum $\propto \omega^s \exp(-s\omega/\omega_0)$ centered at $\omega_0 = ck_0$. The parameters are $k_0 z_R =10$ and $s=20$. The density plots show the intensity distributions $|\psi(x,y=0,z,t)|^2$ at different times, whereas the red dot marks propagation with the subluminal group velocity \eqref{group_velocity_beam}. (b) Numerically calculated retardations of the wavepacket centroid \eqref{wavepacket_centroid}, $Z_c -ct$, for the exact \cite{Vo2024JO} and paraxial \cite{Caron1999JMO} wavepacket solutions with similar parameters, compared with the theoretical prediction based on the group velocity \eqref{group_velocity_beam}.}
\label{fig_packet}
\end{figure}

\subsection{Gaussian beams and wavepackets}
\label{sec:Gaussian}

\blue{We first examine the simplest monochromatic Gaussian beam. Although the intensity distribution of such beam is stationary and its longitudinal position is ill-defined, it can be regarded as the limiting case of very long (quasi-monochromatic) Gaussian wavepackets, which is sufficient to calculate the averaged group velocity in the Fourier domain. The Gaussian beam} is characterized by the ${\bf k}$-space spectrum $\tilde{\psi}(k_\perp) \propto \exp({-k_\perp^2}w_0^2/4)$, where $k_\perp$ is the radial component of the wavevector in cylindrical coordinates, and $w_0 \gg k^{-1}$ is the beam waist.
Substituting this spectrum into Eq.~\eqref{group_velocity}, and employing the paraxial approximation for the longitudinal wavevector component, $k_z = \sqrt{k^2 - k_\perp^2} \simeq k - k_\perp^2/2k$, we calculate the average group velocity:
\begin{equation}
\label{group_velocity_beam}
\langle {v}_{g\,z} \rangle
= c\frac{\int k_z e^{-k_\perp^2w_0^2/2} \,k_\perp d k_\perp}{k \int e^{-k_\perp^2w_0^2/2} \,k_\perp d k_\perp} = c \frac{\langle k_z \rangle}{k}
\simeq c \!\left(1- \frac{1}{2kz_R}\right)\!,
\end{equation}
where $z_R = k w_0^2/2$ is the Rayleigh range, i.e., the characteristic longitudinal scale of beam diffraction. Equation~\eqref{group_velocity_beam} means that a paraxial Gaussian wavepacket accumulates a half-wavelength retardation after propagating a distance $2\pi z_R$, i.e., about six Rayleigh ranges, see Fig.~\ref{fig_packet}. This estimate agrees with Refs. \cite{Giovannini2015, Bareza2016SR, Gouesbet2016}, but the retardation effect calculated in Ref.~\cite{Bliokh2023JPA} is twice as small. This difference arises because Ref.~\cite{Bliokh2023JPA} considered a 2D Gaussian wavepacket in the $(x,z)$ plane, whereas here we treat a fully 3D beam. The number of transverse dimensions is doubled, and so is the resulting correction.   

To verify this result, we employ analytic solutions for diffracting Gaussian-like wavepackets $\psi({\bf r},t)$, both exact \cite{Vo2024JO} and paraxial \cite{Caron1999JMO}. Since these solutions are given in real space-time, we use the natural simplified expression for the wavepacket centroid:
\begin{equation}
\label{wavepacket_centroid}
{\bf R}_{c}
= \frac{\int {\bf r} |\psi|^2 \, d^3 {\bf r}}{\int |\psi|^2 \, d^3 {\bf r}}\,.
\end{equation}
This centroid differs from the energy centroid \eqref{energy_centroid} by the absence of the $\omega$ weighting factors under integrals in the momentum-space representation. 
However, for the paraxial regime considered here and for narrow frequency spectra, this distinction is not essential \cite{Bliokh2023JPA} (see also Section~\ref{sec:wavefunction} below).

\begin{figure}[t]
\centering
\includegraphics[width=\linewidth]{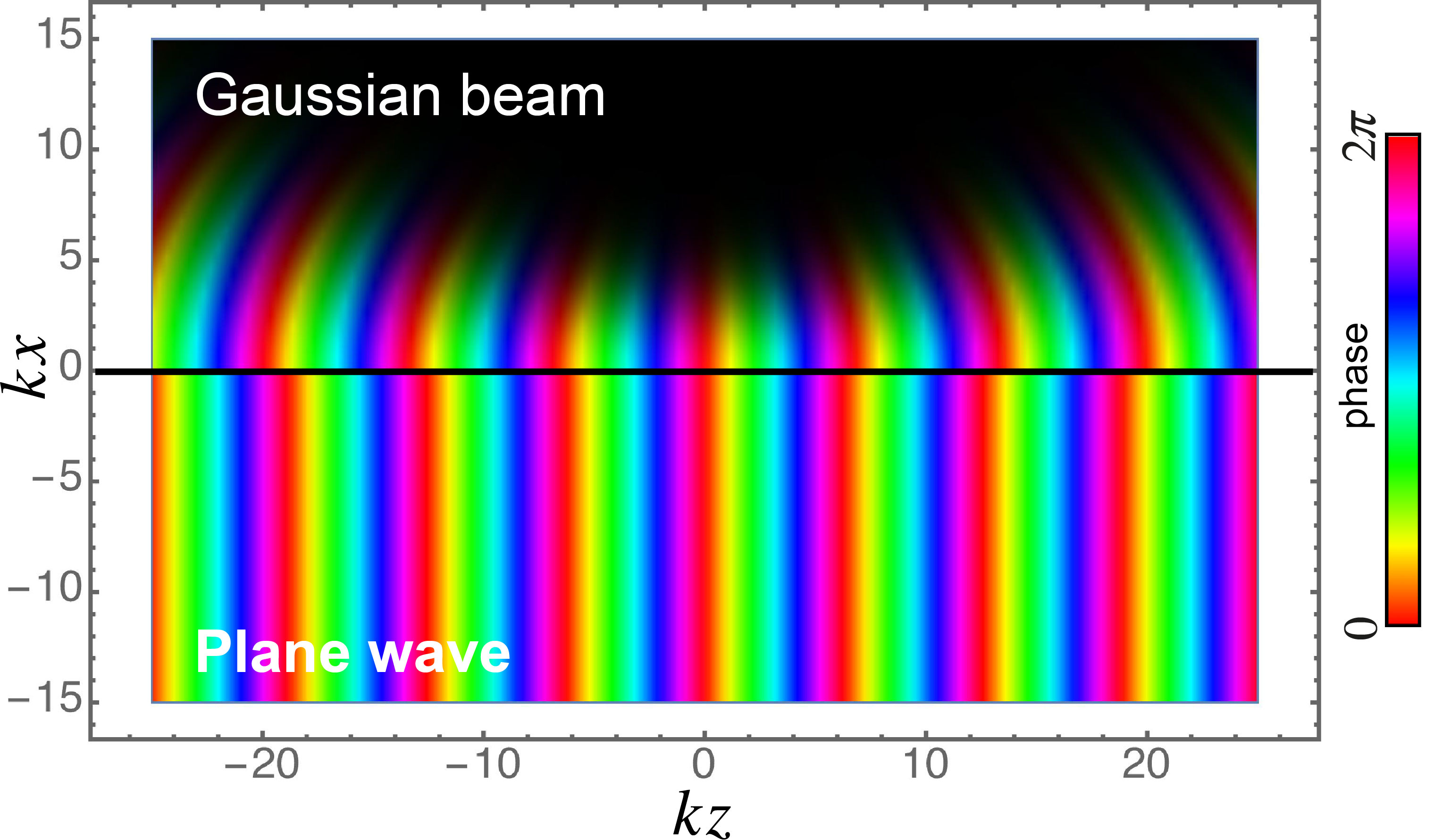}
\caption{Phase distribution in a $z$-propagating Gaussian beam \eqref{Gaussian_beam} with $kw_0 = 5$ (shown in the $x>0$ half-plane) and in a plane wave with the same $k$ (shown in the $x<0$ half-plane). 
The larger spacing between phase fronts in the Gaussian beam corresponds to a superluminal phase velocity, Eq.~\eqref{phase_velocity_beam}.}
\label{fig_beam}
\end{figure}

Figure~\ref{fig_packet}(a) shows the intensity distributions and centroids \eqref{wavepacket_centroid} for the exact diffracting wavepacket \cite{Vo2024JO} propagating over six Rayleigh ranges from the focal plane. Figure~\ref{fig_beam}(b) shows the retardation of the wavepacket centroid relative to propagation at the speed of light, $Z_c - ct$, calculated using the exact \cite{Vo2024JO} and paraxial \cite{Caron1999JMO} solutions, as well as the subluminal group velocity given by Eq.~\eqref{group_velocity_beam}. 
All three approaches are in good agreement, while the small deviations can likely be attributed to higher-order (post-paraxial) corrections.

Next, we analyze the average phase velocity in a Gaussian beam. 
Following a procedure analogous to that leading to Eq.~\eqref{group_velocity_beam}, we obtain $\langle {v}_{ph\,z} \rangle = c k / {\langle k_z \rangle} = c^2 / \langle {v}_{g\,z} \rangle$.
To illustrate how this velocity manifests itself in real space, we consider the well-known real-space representation of a Gaussian beam \cite{Siegman_book} (see Fig.~\ref{fig_beam}):
\begin{align}
\label{Gaussian_beam}
& \psi (r_\perp,z,t)
\propto \frac{w_0}{w(z)}\exp\!\left(-\frac{r_\perp^2}{w^2(z)} \right) \times \nonumber \\
& \exp\!\left[ikz + i\frac{kr_\perp^2 z}{2(z_R^2 + z^2)} - i \arctan\!\left(\frac{z}{z_R} \right) -i\omega t \right]\!.
\end{align}
Here, $r_\perp$ is the radial cylindrical coordinate, and $w(z) = w_0 \sqrt{1+z^2/z_R^2}$ is the $z$-dependent beam radius (equal to $w_0$ at the  waist plane $z=0$). The local phase velocity near the focal plane, $|z|\ll z_R$, can be determined from the phase in the second exponential of Eq.~\ref{Gaussian_beam}. Expanding to leading order in $z/z_R$, we obtain:
\begin{equation}
\label{phase_velocity_local}
\Phi \simeq  \left( k + \frac{kr_\perp^2 }{2z_R^2} - \frac{1}{z_R} \right)\!z -\omega t .
\end{equation}
Averaging this phase over $r_\perp$ with the Gaussian intensity profile, and using 
\[
\langle r_\perp^2 \rangle = \frac{\int r_\perp^2 e^{-2r_\perp^2/w_0^2}\,r_\perp d r_\perp}{\int e^{-2r_\perp^2/w_0^2}\,r_\perp d r_\perp} = \frac{w_0^2}{2}\,,
\]
we derive the average phase velocity:
\begin{equation}
\label{phase_velocity_beam}
\langle v_{ph\, z} \rangle \simeq  \frac{\omega}{k- 1/kw_0^2} \simeq c\left( 1 + \frac{1}{2k z_R} \right) ,
\end{equation}
in agreement with the average group velocity \eqref{group_velocity_beam}. 
The superluminal phase velocity is clearly visible in Fig.~\ref{fig_beam}, which shows that the mean spacing between wavefronts in a Gaussian beam exceeds that of a plane wave with the same frequency. 

It is instructive to relate this analysis to Ref.~\cite{Feng2001OL}, which established a direct connection between the Gouy phase $\Phi_G = - \arctan(z/z_R) \simeq -z/z_R$ and the average longitudinal wavevector component: $\langle k_z \rangle = k +d\Phi_G/dz$. 
In that work, however, the average longitudinal wavevector component was expressed via transverse component as $\langle k_z \rangle = k - \langle k_\perp^2 \rangle/k$, while we used $\langle k_z \rangle = k - \langle k_\perp^2 \rangle/2k$. In our formalism, the average $k_z$ includes contributions from both the Gouy-phase term and the average phase-curvature term: 
$\langle \Phi_R \rangle \simeq {k \langle r_\perp^2 \rangle z}/{2z_R^2} = z/2z_R$. Accordingly,
$\langle k_z \rangle = k +d\Phi_G/dz + d \langle \Phi_R \rangle/dz \simeq k - 1/2z_R$.
The pure Gouy-phase effect is seen in Fig.~\ref{fig_beam} as an increased wavefront spacing on the beam axis, $r=0$. 
However, the average phase velocity requires averaging over the transverse coordinate, which introduces the phase-curvature contribution $\Phi_R$ and reduces the overall deviation $\langle k_z \rangle - k$ by a factor of two compared to the Gouy-phase contribution alone.

\subsection{Higher-order modes}
\label{sec:higher-order}

Next, we consider higher-order free-space modes (e.g., Laguerre-Gaussian or Hermite-Gaussian), which can further enhance the subluminal-group and superluminal-phase effects \cite{Bouchard2016O, Bareza2016SR}. 

For example, the real-space field of Laguerre-Gaussian beams is given by  \cite{Siegman_book}: 
\begin{align}
\label{LG_beam}
& \psi_{LG} \propto 
\frac{w_0}{w}
\!\left(\frac{\sqrt{2}r}{w}\right)^{|\ell|} L_p^{|\ell|}\!\left(\frac{2r^2}{w^2} \right)\! 
\exp\!\left(-\frac{r_\perp^2}{w^2} \right) e^{i\Phi}, \\
& \Phi = kz + \frac{kr_\perp^2 z}{2(z_R^2 + z^2)} - (N+1)\arctan\!\left(\frac{z}{z_R}\right)+\ell\varphi -\omega t,
\nonumber
\end{align}
where $w = w_0 \sqrt{1+z^2/z_R^2}$, $L_p^{|\ell|}$ are generalized Laguerre polynomials, $\ell = 0, \pm 1, \pm 2, ...$ is the azimuthal index, $p = 0, 1,2, ...$ is the radial index, $N= |\ell| + 2p$, and $\varphi$ is the azimuthal angle with respect to the $z$-axis. 
The plane-wave spectrum of these beams reads
\begin{align}
\label{LG_beam_k}
& \tilde{\psi}_{LG} \propto 
\!\left( \frac{w_0k_\perp}{\sqrt{2}} \right)^{|\ell|} 
L_p^{|\ell|}\!\left(\frac{w_0^2k_\perp^2}{2} \right)\!
\exp\!\left(-\frac{k_\perp^2w_0^2}{4}\right)\!e^{i\ell\phi},
\end{align}
where $\phi$ is the azimuthal angle in ${\bf k}$-space with respect to the $k_z$-axis.

Substituting Eq.~\eqref{LG_beam_k} into Eq.~\eqref{group_velocity} and using the integral relations from Ref.\cite{Phillips1983AO}, we derive, analogously to Eq.~\eqref{group_velocity_beam},
\begin{align}
\label{group_velocity_LG}
\langle {v}_{g\,z} \rangle =
c\,\frac{\langle k_z \rangle}{k}\ 
\simeq c \left(1- \frac{N+1}{2kz_R}\right)\!.
\end{align}
Using the phase in Eq.~\eqref{LG_beam}, linearized in $z$ near the focal plane ($|z| \ll z_R$), we obtain, similarly to Eqs.~\eqref{phase_velocity_local}--\eqref{phase_velocity_beam},
\begin{equation}
\label{phase_velocity_LG}
\langle v_{ph\, z} \rangle \simeq  
c\left( 1 + \frac{N+1}{2k z_R} \right) ,
\end{equation}

Equations~\eqref{group_velocity_LG} and \eqref{phase_velocity_LG} show that the subluminal-group and superluminal-phase effects are enhanced in higher-order Laguerre-Gaussian beams by a factor of $(N+1)$. This agrees with the calculations and measurements in \cite{Bareza2016SR, Bouchard2016O}. This enhancement is directly related to the increased difference $k-\langle k_z \rangle \simeq \langle k_\perp^2 \rangle/2k = (N+1)/2z_R$, or, equivalently, to the increased Gouy-phase and phase-front curvature terms with  $k\langle r_\perp^2 \rangle/2z_R^2 = (N+1) /2z_R$ \cite{Phillips1983AO}. 

Thus, for generic paraxial wavepackets or beams, the deviations of the group and phase velocities from $c$ can be expressed as \cite{Giovannini2015, Gouesbet2016, Bareza2016SR, Saari2018PRA}
\begin{equation}
\label{transverse_k}
1 - \frac{\langle v_{g\, z }\rangle}{c}  \simeq \frac{\langle v_{ph\, z} \rangle}{c} -1 \simeq  
\frac{\langle k_\perp^2 \rangle}{2k^2}
\end{equation}
As another example, higher-order Bessel beams \cite{McGloin2005CP} are characterized by a ring-like ${\bf k}$-space spectrum with $k_\perp  = {\rm const}$, $k_z = \sqrt{k^2 - k_\perp^2} = {\rm const}$, and the azimuthal phase factor $e^{i\ell \phi}$. For such beams, Eq.~\eqref{transverse_k} shows that the deviations of the group and phase velocities are determined by $\langle k_\perp^2 \rangle = k_\perp^2$ and are independent of the beam order $\ell$.


\section{Quantum-mechanical approach}

\subsection{Riemann–Silberstein formalism}
\label{sec:RS}

Finally, we address the problem of photon velocities within a quantum-mechanical formalism. 
Maxwell’s equations can be written in the form of a Dirac-like equation for a massless particle, where the complex Riemann–Silberstein vector ${\bf F} = ({\bf E} + i {\bf H})/\sqrt{2}$ (composed of the real electric and magnetic fields) plays the role of a vector wavefunction \cite{BB1996, Bialynicki-Birula2013JPA}:
\begin{equation}
\label{Maxwell}
i \frac{\partial {\bf F}}{\partial t} = c(\hat{\bf S} \cdot {\bf p}) {\bf F} \equiv \hat{\mathcal{H}}\, {\bf F}\,.
\end{equation}
Here, ${\bf p} = -i {\boldsymbol{\nabla}}$ is the momentum operator (we set $\hbar=1$, which does not affect the main result), $\hat{\mathcal{H}}$ is the effective Hamiltonian operator, and 
\begin{equation}
\label{Spin}
\hat{S}_x = \left( \begin{matrix} 0 & 0 & 0 \\
0 & 0 & -i \\ 
0 & i & 0 \end{matrix} \right)\!,~
\hat{S}_y = \left( \begin{matrix} 0 & 0 & i \\
0 & 0 & 0 \\ 
-i & 0 & 0 \end{matrix} \right)\!,~
\hat{S}_z = \left( \begin{matrix} 0 & -i & 0 \\
i & 0 & 0 \\ 
0 & 0 & 0 \end{matrix} \right)
\nonumber
\end{equation}
are the spin-1 operator matrices (generators of SO(3) rotations) \cite{BB1996, Bliokh2015PR}. 

Using the canonical coordinate operator ${\bf r}$, the Heisenberg equation of motion yields
\begin{equation}
\label{EOM}
 {\bf v}\equiv \frac{d {\bf r}}{d t} = [\hat{\mathcal{H}},{\bf r}] = c\, \hat{\bf S}\,,
\end{equation}
where we used the canonical commutation relations $[r_i,p_j] = i \delta_{ij}$, with $\delta_{ij}$ being the Kronecker delta.

It may seem puzzling that the velocity \eqref{EOM} is proportional to the spin operator. However, in the Riemann-Silberstein formalism, this operator is associated with the direction of propagation of plane waves rather than with spin or polarization (similarly to $\hat{\boldsymbol{\alpha}}$-matrices in the Dirac equation \cite{Thaller_Dirac}). Indeed, it is not difficult to show that an arbitrary polarized electromagnetic plane wave propagating along the $z$-axis is described by the Riemann-Silberstein vector ${\bf F} \propto (1,i,0)^T$, which is an eigenvector of $\hat{S}_z$ with eigenvalue 1. Therefore, the corresponding velocity of this plane wave is equal to $c$ and directed along the $z$-axis. Moreover, the local expectation value ${\bf F}^* \cdot (\hat{\bf S}) {\bf F} = {\rm Im} ({\bf F}^* \times {\bf F}) = ({\bf E} \times {\bf H}) =c {\bf P}$ is proportional to the electromagnetic momentum density. Accordingly, the normalized expectation value of the velocity \eqref{EOM} becomes
\begin{equation}
\label{velocity_quantum}
\langle {\bf v} \rangle = c\,\langle \hat{\bf S}\rangle
\equiv c\, \frac{\langle {\bf F}|\hat{\bf S}|{\bf F}\rangle}{\langle {\bf F}|{\bf F}\rangle} 
= c\, \frac{2\int ({\bf E} \times {\bf H})\, d^3 {\bf r}}{\int ({\bf E}^2 + {\bf H}^2)\, d^3 {\bf r}}\,,
\end{equation}
where the bra-ket inner product implies spatial integration, and we used ${\bf F}^*\cdot {\bf F} \equiv |{\bf F}|^2 = ({\bf E}^2 + {\bf H}^2)/2 =U$.
Thus, Eq.~\eqref{velocity_quantum} is fully equivalent to Eq.~\eqref{energy_velocity} for the energy-centroid velocity: $\langle {\bf v} \rangle = {\bf v}_E$.

It is instructive to examine the relation of the photon velocity \eqref{EOM} and \eqref{velocity_quantum} to the long-standing problem of the {\it photon position operator} \cite{Bacry_book}. The problem arises because the canonical position operator ${\bf r}$ is inconsistent with the transversality condition imposed by Maxwell’s equations, $\boldsymbol{\nabla}\cdot {\bf E} = \boldsymbol{\nabla}\cdot {\bf H} = 0$, i.e., ${\bf p}\cdot {\bf F} = 0$ (momentum-space translations generated by ${\bf r}$ violate the transversality of a generic transversal field). 
To resolve this issue, one can consider an alternative position operator projected onto the subspace of transverse (i.e., divergenceless) fields \cite{Bacry_book, Bliokh2010PRA, Bliokh2017PRA}: ${\bf r}' = {\bf r} + ({\bf p} \times \hat{\bf S})/p^2$.
Importantly, for physically admissible (transverse) fields, this operator yields the same expectation values as the canonical operator ${\bf r}$. The corresponding equation of motion becomes
\begin{equation}
\label{EOM_P}
 {\bf v}' \equiv \frac{d {\bf r}'}{d t} = [\hat{\mathcal{H}},{\bf r}'] = c\, \frac{{\bf p}({\bf p}\cdot\hat{\bf S})}{p^2}
 = \frac{{\bf p} \hat{\mathcal{H}}}{p^2}\,,
\end{equation}
where we used the SO(3) commutation relations $[\hat{S}_i,\hat{S}_j] = i \epsilon_{ijk} \hat{S}_k$ ($\epsilon_{ijk}$ is the Levi-Civita symbol).
Notably, the velocity \eqref{EOM_P} is explicitly aligned with the canonical momentum ${\bf p}$. 
It is easy to see that the projections of the velocities \eqref{EOM} and \eqref{EOM_P} onto the momentum coincide: ${\bf v} \cdot {\bf p} = {\bf v}' \cdot {\bf p}$. Hence, the expectation value $\langle {\bf v}' \rangle$ must coincide with Eq.~\eqref{velocity_quantum}, i.e., the energy-centroid velocity. The difference between the two velocities has the form of a curl-like field orthogonal to the momentum, ${\bf v}' - {\bf v} = c\,{\bf p}\times ({\bf p}\times \hat{\bf S})/p^2$, which vanishes upon integration. 

Let us now consider the time derivative of the product ${\bf r}\cdot {\bf p} = {\bf r}'\cdot {\bf p}$. Since $d{\bf p}/dt = [\hat{\mathcal{H}},{\bf p}] = {\bf 0}$, we obtain: 
\begin{equation}
\label{EOM_2}
\frac{d ({\bf r}\cdot {\bf p})}{d t} = {\bf v}\cdot {\bf p} = 
\hat{\mathcal{H}}\,.
\end{equation}
This relation resembles the quantum-mechanical analogue of Eq.~\eqref{momentum_centroid}. Introducing a `momentum velocity' via relation ${\bf v}_{P1} \cdot \langle {\bf p}\rangle = \langle {\bf v}\cdot {\bf p} \rangle$, and using Eq.~\eqref{EOM_2}, we have
\begin{equation}
\label{momentum_velocity_quantum_1}
{\bf v}_{P1} \cdot \langle {\bf p}\rangle = \langle \hat{\mathcal{H}} \rangle\,,
\end{equation}
which formally resembles a quantum-mechanical version of Eq.~\eqref{momentum_velocity}.

However, the proper expression of Eq.~\eqref{momentum_velocity} in the Riemann-Silberstein formalism is:
\begin{equation}
\label{momentum_velocity_quantum_2}
{\bf v}_P \cdot \langle {\bf F}|\hat{\bf S}| {\bf F}\rangle = c\, \langle {\bf F}| {\bf F} \rangle \,,~~{\rm i.e.},~~{\bf v}_P \cdot \langle \hat{\bf S}\rangle = c\,.
\end{equation}
Combined with Eq.~\eqref{velocity_quantum}, this guaranties ${\bf v}_P\cdot \langle {\bf v} \rangle = c^2$, and is not equivalent to Eq.~\eqref{momentum_velocity_quantum_1}.
The origin of the discrepancy between Eqs.~\eqref{momentum_velocity_quantum_1} and \eqref{momentum_velocity_quantum_2} is considered below.

\subsection{Peculiarities related to the photon wavefunction}
\label{sec:wavefunction}

Let us examine expressions \eqref{velocity_quantum}, \eqref{momentum_velocity_quantum_1}, and \eqref{momentum_velocity_quantum_2} in the momentum representation. Denoting the plane-wave Fourier components of the Riemann-Silberstein vector by $\tilde{\bf F}({\bf k}) e^{-i\omega({\bf k}) t} = (\tilde{\bf E} + i \tilde{\bf H}) e^{-i\omega t} /\sqrt{2}$, and noting that in this representation the momentum and Hamiltonian operators become ${\bf p} \to {\bf k}$ and $\hat{\mathcal{H}} \to \omega$, we obtain:
\begin{align}
\label{velocity_quantum_k}
\langle {\bf v} \rangle & = {\bf v}_E
 = c\, \frac{\int ({\bf k}/k)\, |\tilde{\bf F}|^2\, d^3 {\bf k}}{\int |\tilde{\bf F}|^2\, d^3 {\bf k}}\,, \\
\label{momentum_velocity_quantum_1k}
{\bf v}_{P1} & \cdot \int {\bf k}\, |\tilde{\bf F}|^2\, d^3 {\bf k}
 = {\int \omega\,|\tilde{\bf F}|^2\, d^3 {\bf k}}\,, \\
\label{momentum_velocity_quantum_2k}
{\bf v}_{P} & \cdot \int ({\bf k}/k)\, |\tilde{\bf F}|^2\, d^3 {\bf k}
 = c\,{\int |\tilde{\bf F}|^2\, d^3 {\bf k}}\,.
\end{align}
Here, we used the relations between plane-wave amplitudes following from Maxwell's equations: $|\tilde{\bf F}|^2 = |\tilde{\bf E}|^2= |\tilde{\bf H}|^2$, ${\bf k}\cdot \tilde{\bf F} = 0$, and $\tilde{\bf H} = ({\bf k}/k) \times \tilde{\bf E}$, which yield $\tilde{\bf E} \times \tilde{\bf H} = ({\bf k}/k) |\tilde{\bf E}|^2$. The expectation value of velocity \eqref{EOM_P}, $\langle {\bf v}' \rangle$, is also given by Eq.~\eqref{velocity_quantum_k}, because the momentum and Hamiltonian operators become in the momentum representation ${\bf k}$ and $\omega$, respectively. 

Although Eqs.~\eqref{velocity_quantum}, \eqref{velocity_quantum_k} and \eqref{momentum_velocity_quantum_2}, \eqref{momentum_velocity_quantum_2k} are equivalent to electromagnetic field-theory Eqs.~\eqref{energy_velocity} and \eqref{momentum_velocity}, their structure differs from the wave-packet relations \eqref{group_velocity} and \eqref{phase_velocity}. Specifically, they involve the average of the ratio $\langle {\bf k}/k \rangle$, while the group and phase velocities in Eqs.~\eqref{group_velocity} and \eqref{phase_velocity} are expressed through the ratio of averages, $\langle {\bf k} \rangle / \langle k \rangle = c\, \langle {\bf k} \rangle / \langle \omega \rangle$. In general, these quantities are not identical. 

The resolution of this apparent discrepancy lies in the fact that the Riemann-Silberstein vector is not actually a photon wavefunction \cite{BB1996, Bialynicki-Birula2013JPA}. Indeed, its intensity $|{\bf F}|^2$ is proportional to the {\it energy density}, not to the {\it probability density} of photons. (That is why by calculating the expectation value of the velocity in the Riemann-Silberstein formalism, we obtained the energy-centroid velocity \eqref{velocity_quantum}.) In fact, the photon wavefunction and probability density cannot be defined in real space due to the Weinberg--Witten theorem \cite{Weinberg1980PLB}. Nonetheless, one can define a photon wavefunction in momentum space as $\tilde{\boldsymbol{\psi}} = \tilde{\bf F}/\sqrt{\omega}$, so that the energy operator becomes $\omega$ in this representation: $\int |{\bf F}|^2 d^3 {\bf r} = \int \omega |\tilde{\boldsymbol{\psi}}|^2 d^3{\bf k}$. Making the substitution $\tilde{\bf F} \to \sqrt{\omega}\, \tilde{\boldsymbol{\psi}}$ in Eqs.~\eqref{velocity_quantum_k} and \eqref{momentum_velocity_quantum_2k}, these expressions become entirely similar to wavepacket Eqs.~\eqref{group_velocity} and \eqref{phase_velocity}. 

By contrast, Eq.~\eqref{momentum_velocity_quantum_1k} for velocity ${\bf v}_{P1}$ acquires additional factors of $\omega$ under integrals, which lack a clear physical interpretation. This occurs because Eq.~\eqref{momentum_velocity_quantum_1} determines the expectation values of the canonical momentum and effective-Hamiltonian operators using the Riemann-Silberstein vector rather than the proper photon wavefunction. Consequently, this alternative definition of `momentum velocity' is not physically meaningful.  

Finally, we note that the wavepacket centroid \eqref{wavepacket_centroid}, which can be interpreted as the {\it probability} centroid for the wavefunction $\psi({\bf r},t)$, in momentum representation takes the form 
\begin{equation}
\label{wavepacket_centroid_k}
{\bf R}_c
= \frac{\int \, \tilde{\psi}^* e^{i\omega t} (i{\boldsymbol{\nabla}}_{\bf k}) \tilde{\psi}e^{-i\omega t}\,d^3{\bf k}}{\int |\tilde{\psi}|^2  \,d^3{\bf k}}\,.
\end{equation}
In general, this centroid differs from the energy centroid \eqref{energy_centroid}. This difference can be important, for example, in calculations of the transverse angular momentum of vortex wavepackets \cite{Bliokh2012PRL, Bliokh2023} and in other subtle problems \cite{Bliokh2025PLA}. However, for the group velocity of a paraxial Gaussian-like wavepacket, both the energy and probability centroids yield essentially the same subluminal result \cite{Bliokh2023JPA}, the difference appearing only at higher order in the small parameter.

Thus, precise definitions of the energy centroid and photon wavefunction are essential for formal consistency. In practice, however, for measurements of photon group or phase velocities, it is typically sufficient to treat the electric field as an effective real-space wavefunction and employ the wave-packet analysis presented in Section~\ref{sec:wavepacket}.



\section{Conclusions}

In summary, we have analyzed the average group and phase velocities of electromagnetic wavepackets propagating in free space. 
We have shown that spatial confinement in the transverse direction inevitably leads to a subluminal group velocity and a superluminal phase velocity, whose product equals $c^2$. 
These velocities can also be identified as the energy-centroid velocity and `momentum velocity', respectively.
\blue{An important relativistic consequence of the subluminal group velocity is that any localized photon state admits a well-defined {\it rest frame}, where its mean momentum vanishes, and it becomes a superposition of counter-propagating (i.e., standing) waves \cite{Bliokh2026}.}

To elucidate the origin and consistency of this result, we examined the problem within three complementary frameworks: electromagnetic field theory, scalar wavepacket evolution, and the quantum-mechanical formalism. These approaches yield mutually consistent conclusions, although each framework highlights a distinct aspect of the underlying physics: field-theoretical conservation laws, wavepacket diffraction effects, and the role of proper photon-wavefunction definition in quantum mechanics.

Overall, our results demonstrate that the subluminal group velocity and superluminal phase velocity of photons are not paradoxical but represent fundamental and internally consistent features of wave propagation in free space.


\begin{acknowledgements}
I am grateful to Sergey N. Shevchenko for inviting me to contribute this article to the special issue of {\it Low Temperature Physics} dedicated to the Kharkiv Quantum Seminar, organized in Ukraine during the wartime. 
I also acknowledge fruitful correspondence with Titouan Gadeyne, who drew my attention to the `momentum velocity' concept and textbook \cite{Schwinger_book}, as well as with Miguel A. Alonso, who provided closed-form wavepacket solutions \cite{Vo2024JO}. 
This work was supported by Marie Sk\l{}odowska-Curie COFUND Programme of the European Commission (project HORIZON-MSCA-2022-COFUND-101126600-SmartBRAIN3). 
\end{acknowledgements}

\bibliography{References}

\end{document}